\documentstyle[aps,prb,multicol,epsfig]{revtex}

\def\top#1{\vskip #1\begin{picture}(290,80)(80,500)\thinlines \put(
65,500){\line( 1, 0){255}}\put(320,500){\line( 0, 1){5}}\end{picture}}
\def\bottom#1{\vskip #1\begin{picture}(290,80)(80,500)\thinlines \put(
330,500){\line( 1, 0){255}}\put(330,500){\line( 0, -1){5}}\end{picture}}

\def \beq{\begin{equation}} 
\def \eeq{\end{equation}} 
\def \barr{\begin{eqnarray}}
\def \earr{\end{eqnarray}}
\def \bmul{\begin{multicols}{2}}
\def \emul{\end{multicols}}
\def \s{\sigma^{3D}_z}
\def \br{{\bf r}} 
\def \bR{{\bf R}}
\def \bv{{\bf v}}
\def \bV{{\bf V}}
\def \la{\langle}
\def \ra{\rangle}
\def \rar{\rightarrow}
\def \df{\delta \phi}
\def \nf{\nabla \phi_v}
\def \D{\Delta}
\def \d{\delta}
\def \pv{\phi_v}
\def \pd{\partial}
\def \jx{j_x}
\def \jz{J_z}
\def \rn{R_x}

\def \rz{\rho_z}
\def \r0{\rho_0}
\def \r1{\rho_1}
\def \rqp{\rho_{qp}}
\def \rc{R_{int}}
\def \jc{j_c}

\def \js{J_{out}}
\def \j0{J_0}
\def \ji{J_{in}}
\def \di{d_{int}}
\def \rco{R_{int, 0}}
\def \td{T_d}
\def \r{\rho}
\def \rd{\rho^{3D}}
\def \a{\alpha}
\def \uz{\bf{\hat{z}}}
\def \uy{\bf{\hat{y}}}
\def \fd{f^{3D}}
\def \jd{j^{3D}}
\def \ed{E^{3D}}
\def \jr{j^{red}}
\def \er{E^{red}}
\def \a{\alpha}
\def \lxy{L_{xy}}
\def \x{\kappa}
\title{Vortex shear effects in layered superconductors}
\author{V.~Braude and~A.~Stern}
\address{Department of Condensed Matter Physics,  The Weizmann Institute of 
Science, Rehovot 76100, Israel}
\date{\today}

\begin{document}

\maketitle
\begin{abstract}
Motivated by recent transport and magnetization measurements in BSSCO
samples [B.~Khaykovich 
{\it et al.}, Phys.~Rev~B {\bf 61}, R9261 (2000)], 
we present a simple macroscopic model describing effects of inhomogeneous
current distribution and shear in a layered superconductor. Parameters
of the model are deduced from a microscopic calculation. Our model
accounts for the strong current non-linearities and the re-entrant temperature
dependence observed in the experiment.


\end{abstract}
\begin{multicols}{2}
\section{Introduction}

Transport measurements are  widely used in studies of vortex dynamics of
high-$T_c$ superconductors. When the current distribution in the sample
is not homogeneous, the results of the measurements are usually interpreted
 in terms of a
local resistivity tensor.
Due to high anisotropy of these materials
the in-plane resistivity
$\r_{xy}$ is much smaller than the out-of-plane resistivity $\r_z$. Commonly 
the resistivity
is assumed to be a local function of the current density, and to depend on 
the applied magnetic field and the temperature \cite
{busch,safar,cho}. 
A recent experiment by Khaykovich {\it et al.} \cite{khay} does not
fit into this scheme.
In this experiment 
transport and magnetization measurements 
\begin{figure}[hbpt]
 \centerline{\epsfig{figure=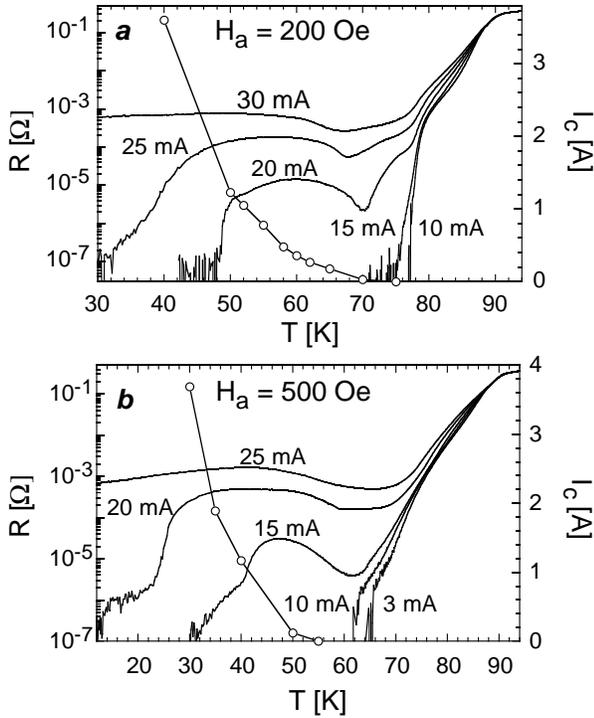,width=8.cm}}
\vspace{0.2cm}
\caption{Resistance at various $I_a$ (left axis, log scale) and magnetically
  measured critical current (right axis, linear scale, open circles) vs T
  for the irradiated sample, $H_a=200$ Oe (a) and $H_a=500$ Oe (b) (taken from
Khaykovich \it{et al.}~\protect\cite{khay})}
\label{expt}
\end{figure}
\noindent in BSCCO
crystals at elevated transport currents and perpendicular magnetic field
are performed, using high quality BSCCO platelets
with current leads attached to the top surface and an array of
2DEG Hall sensors to the  bottom surface. At a first glance, the pictures 
that emerge from the transport
and the magnetization measurements are mutually contradicting. 
Transport measurements
reveal finite resistivity below the magnetic irreversibility line, in the 
superconducting state.
This 
resistivity is non-monotonic with temperature, showing re-entrant behavior,
and non-linear with current. 
As seen in the graphs of $R$ vs. $T$, Fig. \ref{expt}, at low transport currents $R(T)$ is monotonic, dropping below 
experimental resolution when temperature is reduced. At elevated currents, the resistance initially drops as 
$T$ is lowered, but then goes up, the bump being steeper at lower currents.
Also $R(T)$ shows strong non-linearity, so that an increase of the current
by $30 \%$ or less may result in enhancement of $R$ by orders of magnitude. 
The source of this resistance is, presumably, vortex flow as a response to the electric current. 

\begin{figure}[hbpt]
 \centerline{\epsfig{figure=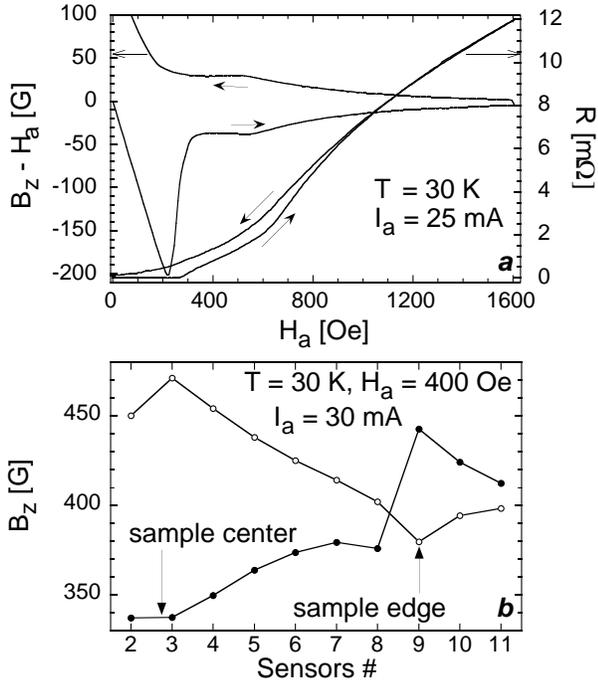,width=8.cm}}
 \caption{(a) Resistance (right axis) and hysteretic magnetization loop in 
the sample center (left axis) vs. $H_a$ at $T$=30 K and $I_a$=25 mA. (b) 
Profile of magnetic induction across the sample at 400 Oe on increasing
($\bullet$) and decreasing ($\circ$) fields (taken from Khaykovich 
\it{et al.}~\protect\cite{khay}).}
\label{magn}
\end{figure}

In contrast, local magnetization measurements in
the presence of transport current, shown in Fig. \ref{magn}a, indicate that the vortices are pinned. 
These measurements can be well described in terms of the Bean model of the 
critical
state \cite{bean,brandt}. The model states that below the irreversibility 
line the local
current density equals either zero, or the critical current density, directed
in such a way as to obtain the total transport current and the magnetization.
The spatial distribution of the magnetic field is then given by the
Biot-Savart law \cite{zeldov}.
Since the current density nowhere exceeds the critical one, the Bean model
predicts zero resistance. 
Within the 
Bean model finite resistivity can be 
expected only
above the magnetically measured irreversibility line, which in Fig.
 \ref{magn}a occurs above
1600 Oe. Indeed, at low $I_a$ the measurements (carried below the irreversibility line) show practically zero resistance. However, at 
elevated currents, substantial resistance is measured concurrently with
the hysteretic magnetization well below the irreversibility line, as seen in Fig. \ref{magn}a.
Figure \ref{magn}b shows the corresponding field profile $B_z(x)$,
obtained by the array of Hall sensors at 400 Oe in presence of transport
current on increasing and decreasing $H_a$. A clear Bean profile is observed.
Fitting this profile to the 

\noindent theoretical field distribution in platelet sample
results in total critical current of $I_c=4.2$ A, which is more than two 
orders of magnitude higher than the transport current of 25 mA. Figure 
\ref{expt} shows $I_c(T)$ determined from the Bean profiles together with
the resistive data. The re-entrant resistance always occurs in the region 
where zero resistance is expected, since the transport current is much 
lower than the critical current.

Thus, the main puzzling observations of Khaykovich {\it et al.} 
are the nonvanishing resistance 
below the irreversibility line, which indicates flux flow, coexisting with
magnetization measurements which indicate that the vortices are pinned, 
the re-entrant behavior of
the resistance with the temperature and its strongly nonlinear dependence
on the current.

Khaykovich  {\it et al.} \cite{khay} suggest the following qualitative 
understanding of the observation. BSCCO,
being a strongly anisotropic type II high $T_c$ superconducting material,
consists of superconducting $\mbox{CuO}_2$ layers, separated by insulating 
barriers. Each layer can carry current, resulting in total parallel
current along the sample. Also, due to Josephson coupling between
the layers, current can flow perpendicular to the layers. Because of
large anisotropy a typical 
ratio of the perpendicular and parallel
resistivities is $\simeq 10^4$ in the normal state. In perpendicular
magnetic field the flux penetrates the system in form of vortices, but,
due to weak interlayer coupling, these are two dimensional ``pancakes'', 
rather
than three dimensional filaments. Pancakes in the same layer repel
one another, while those in different layers attract via Josephson and
magnetic coupling \cite{brandt}.
 In the experiment, the leads are attached to the top surface
of the crystal. Hence the current distribution is non-homogeneous along
the sample thickness, planes near the bottom of the crystal carrying
much lower current than those at the top. As temperature decreases, 
pinning of vortices becomes more effective. Eventually the critical current
density exceeds current density near the bottom. Then pancake vortices
at the bottom stop moving, while pancakes at the top maintain their high
velocity, since current density there is much higher than the critical 
current density. As a result, velocity gradient of pancake motion between
different layers is increased. This, in turn, leads to shear-induced
phase slippage between the adjacent $\mbox{CuO}_2$ planes, reducing the
Josephson coupling and increasing the perpendicular resistance $\rz$.    
The larger $\rz$ causes the current to flow in a thinner part of the sample,
thus
making the process self-enhancing. Since all of the transport current 
flows in a few layers near the  top of the sample, finite resistance
exists at currents much lower than the critical current expected from the
Bean model. Magnetization measurements, on the other hand, measure the magnetic response
of all layers. When the vortices are pinned in most layers, this response
is irreversible. 

In this work we take this qualitative explanation as a starting point and construct 
macroscopic and microscopic models to analyze the experiment. We start by presenting a macroscopic model in
which the sample is assumed to be constructed of a resistive part, an interface and 
a dissipationless part. The perpendicular resistivity of the resistive part is assumed to depend on
"vortex shear". The parameters of this model are introduced phenomenologically. 
We then examine the dependence of the sample's resistance on these parameters, and the conclusions that
may be drawn regarding the dependence of the resistance on the temperature and
current. Following that we construct a microscopic model aimed at deriving an expression relating
the conductivity in the direction perpendicular to the layers to the inter-layer variation 
of the current parallel to the layers. Finally we compare the conclusions of 
our model to the experimental
findings. Although we find a general agreement, we also point out some 
remaining difficulties, associated 
mostly with the lack of quantitative information regarding  several of the
 parameters of the model.

\section{The macroscopic model}
As we focus here on the consequences of inhomogeneity in the current 
distribution in  $z$ direction,
we use a one-dimensional model in which all quantities can vary 
only in this direction. Since scales of interest are much larger than the
microscopic scale defined by the spacing between adjacent superconducting 
layers, we take a continuous limit in $z$ direction.
\begin{figure}[hbpt]
   \centerline{\psfig{figure=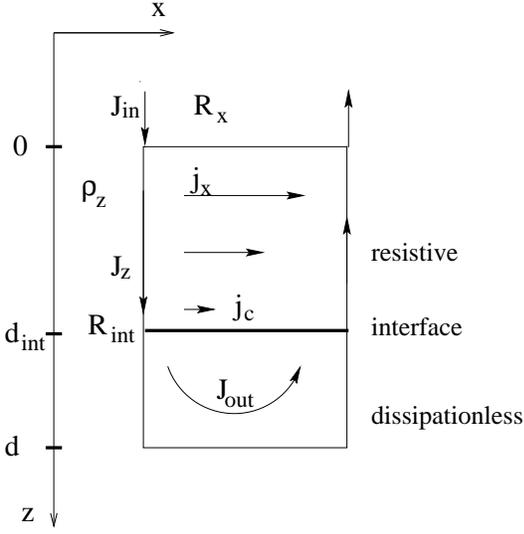,width=7cm}}
   \caption{The macroscopic model.}
   \label{system}
 \end{figure}
The model is described as follows. A current $J_{in}$ is injected into 
a system of depth $d$ from above.
Part of the current then flows horizontally as $\jx$ and the rest - 
vertically down
as $\jz$ (then, of course, returning vertically up at the other end of the
system). Below the depinning temperature $T_d$, when there is non-zero critical current density
$\jc$, the system can be divided into two parts, by the value of the in-plane current $\jx$.
 The upper part of the system 
carries current density larger than
$\jc$, so it has finite resistance, while at
the lower part the current density is smaller than $\jc$, and thus it has
zero resistance. Accordingly, we consider the system as consisting of two
phases: a resistive phase at the top, having parallel resistivity $\rn$ and
perpendicular resistivity $\rz/2$, and a dissipationless phase with zero
resistivity. Note, that since the current first flows down and then up,
the total perpendicular resistivity it experiences is $\rz$. 
Furthermore, we assume that current crossing the interface between the 
two phases faces a resistance $\rc/2$. The position of the interface is
determined by the condition $\jx=\jc$. This condition also fixes
the current $\js$ flowing through the dissipationless region:
\beq \label{eq:jout}
  \js \rc=\jc \rn.
\eeq
At high temperatures $\jc$ is zero, and the system consists only of the
dissipative phase. 
 
The basic equations governing the distribution 
of the current in the dissipative phase are the two Kirchoff equations. 
The continuity 
equation is 
(note that
in the geometry we consider $\jz$ and $\jx$ have different dimensions, since $\jz$ is a two dimensional 
current density,while $\jx$ is a three dimensional 
current density):
\beq \label{eq:cont}
  \partial_z \jz+\jx=0
\eeq
and the equation giving the total voltage is:
\beq \label{eq:volt}
  V=\int_0^z \jz(z') \rz(z') dz'+\jx(z) \rn.
\eeq
As we show below in the microscopic analysis, the $z$-axis resistivity depends
on the difference between $j_x$ in adjacent layers $\pd_z j_x$, and this
dependence may be approximated by 
\beq \label{eq:rhoz}
  \rz=\r_0+\sqrt{\r_1^2+(f \partial_z \jx)^2}=\r_0+\sqrt{\r_1^2+(f \pd_z^2 
  \jz)^2},
\eeq
while $\rn$ is assumed to be a constant parameter. The term $f\partial_z \jx$ in the 
resistivity $\rz$ is a contribution of the "shear" between vortices in different layers to the out-of-plane 
resistance. It originates from the effect of a velocity gradient between vortices in adjacent planes 
on the Josephson coupling between the planes.

Substituting Eq. (\ref{eq:cont}) into Eq. (\ref{eq:volt}) and differentiating 
with respect to $z$ we obtain:
\beq \label{eq:currdistr}
  \jz (\r_0+\sqrt{\r_1^2+(f \partial_z^2 \jz)^2})-\rn \partial_z^2 \jz=0.
\eeq
This equation can be solved only if the condition
\beq \label{eq:cond}
  \jz<\frac{\rn}{\r_0} \pd_z^2 \jz
\eeq
is satisfied. Designating $J\equiv \jz$, $J''\equiv \pd_z^2 \jz$ and solving 
for $J''$, we obtain
\barr \label{eq:secder}
  J''&=&\frac{1}{(\rn/J)^2-f^2} 
 \nonumber \\ &&
\times \left(\rn \r_0/J\pm \sqrt{(\rn \r_1/J)^2+f^2
  (\r_0^2-\r_1^2)}\right).
\earr
The condition (\ref{eq:cond}) requires that plus sign be taken in 
Eq. (\ref{eq:secder}) and that
$J<J_0\equiv \rn/f$. This means that  as $\ji \rar \j0$, both $J''$ and 
$J'$ diverge, so that the voltage $V$ also diverges, and the system 
becomes insulating. In fact, as 
$\ji\rar J_0$, current gradients in the system become large, and then the
quasi-particle channel for $z$-axis currents needs to be taken into account, 
as analyzed below.
When doing this, we find that $J_0$ is actually not a cutoff value for 
the injected
current, but rather a parameter that signifies the importance of shear 
effects.
Thus, when $\ji$ becomes comparable with $J_0$, shear becomes strong, and
the resistance is strongly non-linear
with $\ji$.

\begin{figure}[hbpt]
  \centerline{\psfig{figure=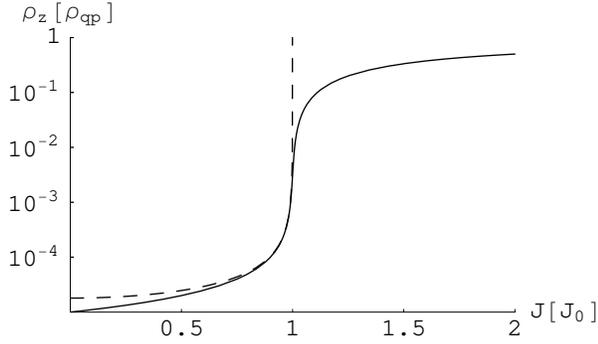, width=8.5cm}}
  \caption{Perpendicular resistivity $\r_z$ for the model without 
  quasiparticle channel, Eq. (\ref{eq:rho1}) (dashed line) and with 
quasiparticle channel, Eq. (\ref{eq:rho2}) (solid line).}
  \label{rho}
\end{figure}

Substituting the solution for $J''$ into Eq. (\ref{eq:rhoz}), the 
perpendicular resistivity can be expressed 
in terms of $J$:
\beq \label{eq:rho1}
  \rz=\r_0\frac{1+\sqrt{r^2+\x^2(1-r^2)}}{1-\x^2},
\eeq
where we used reduced quantities $\x\equiv J/\j0$ and $r\equiv \r_1/\r_0$. It
is plotted in Fig. \ref{rho}.
Again, this is valid for $J$ not too close to $\j0$.

It is possible to integrate Eq. (\ref{eq:secder}). Some intuition to it may 
be obtained by noticing that Eq. (\ref{eq:secder}) may be viewed as an 
equation of motion for a particle whose one dimensional coordinate is $J$,
its "time" is $z$, and the potential 

\begin{figure}
   \centerline{\psfig{figure=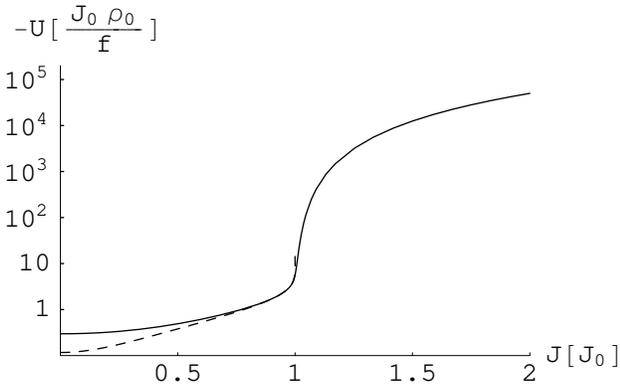,width=8.5cm}}
   \caption{Potential $U(J)$ (taken with minus sign) for the model without
quasiparticle channel, Eq. (\ref{eq:pot1}) (dashed line) and with 
quasiparticle channel, Eq. (\ref{eq:pot2}) (solid line).} 
   \label{potent}
\end{figure}

\noindent it moves in is
\barr \label{eq:pot1}
   U(J)=\frac{\j0 \r_0}{f} \Big(&&\sqrt{r^2+\x^2(1-r^2)}
\nonumber \\ &&
+\ln|1-
   \sqrt{r^2+\x^2(1-r^2)}|  \Big),
\earr
This potential is plotted in Fig. \ref{potent}.

The analysis leading to Eq. (\ref{eq:secder}) neglects inter-layer current flow by means of quasi-particle tunneling. When the current gradient $J''$ gets large, $\rz$ becomes large, and a large portion of the current 
flows perpendicularly
in the form of quasiparticles.
Hence in this
high-gradient limit the perpendicular resistivity should be modeled by two
resistors in parallel. Also, since in this regime the current gradients are
large, a linearized expression for the Josephson channel resistivity, Eq. 
(\ref{eq:rhoz}),
can be used. Thus the Josephson channel carries a resistivity
$\r_0+f\pd_z j_x$, while the quasiparticle channel's resistivity is
$\rqp$. The total perpendicular resistivity is
\beq \label{eq:highres}
  \r_z^{-1}(\pd_z j_x)=(\r_0+f\pd_z j_x)^{-1}+\rqp^{-1}.
\eeq
It is assumed, of course, that $\rqp \gg\r_0$. 
Using this assumption and solving again for the current distribution, 

\noindent we get
a differential equation
\beq \label{eq:secder2}
  J''=\frac{\rqp}{2 f} \left(\x-1 \pm \sqrt{(\x-1)^2+4\x \r_0/\rqp} \right),
\eeq
where again $\x \equiv J/\j0$. In order to have $J''>0$, we need to choose
the plus sign. The perpendicular resistivity can be expressed in terms of $J$:
\beq \label{eq:rho2}
 \r_z=\rqp \frac{\x-1+\sqrt{(\x-1)^2+4\x r_2}}{2\x},
\eeq
where $r_2 \equiv \r_0/ \rqp$. It is plotted in Fig. \ref{rho}.

Integrating Eq. (\ref{eq:secder2}),  we obtain the corresponding 
``potential'', plotted in Fig. \ref{potent}:
\begin{eqnarray} \label{eq:pot2}
 && U(J)=-\frac{\j0 \rqp}{4 f} 
\nonumber \\ && \quad       \times
\Bigg[ (\x-1)^2+4r_2(1-r_2)
{\rm arcsh}{\frac{\x-1+2r_2}{2\sqrt{r_2(1-r_2)}}} \nonumber \\
 && \quad \,\,\,   + (\x-1+2r_2)\sqrt{4r_2
(1-r_2)+(\x-1+2r_2)^2} \Bigg].
\end{eqnarray}
In both cases we may use $U(J)$ together with the boundary conditions
to determine the resistance of the system.
The ``velocity'' of the particle is given by 
\beq \label{eq:jx}
  \partial_z \jz = -\sqrt{2[C-U(J)]}.
\eeq
Here $C$ is a constant determined by the boundary conditions, which require
$\jz(z=0)=\ji$, and either $\jz(d)=0$, in the case where the whole sample
is resistive so that $\jx>\jc$; or $\jz(\di)=\js$ and 
$\partial_z \jz(\di)=\jc$, for the case where the lower part of the sample
is dissipationless and is separated from the upper part by an interface at
depth $\di$. 
The first case takes place at temperatures above $\td$, where $\jc=0$,
while the second case - when the temperature is below $\td$.  
For the second case the derivative $J'(z=0)$ can be found:
\beq
  J'(0)=-\sqrt{2(U(\js)-U(\ji))+\jc^2}.
\eeq 
Finally, after solving for the current profile $J(z)$ we may calculate the resistance of the sample to be 
\beq \label{eq:resist}
  R=\frac{V}{\ji}=-\frac{\rn J'(0)}{J(0)}.
\eeq
We now summarize how the resistance depends on various parameters of the model.

\subsection{Dependence on the injected current $\ji$}
For high temperatures, when the whole sample is 
dissipative, the resistance
increases monotonously with $\ji$, experiencing a sharp increase around $\j0$.
 This
is because larger currents produce larger current gradients, which,
in turn, increase the vertical resistivity $\rz$.

The low temperature case, where there is an interface,  is more 
complicated
and depends on the value of the interface resistance $\rc$. To investigate 
the dependence 
of the resistance on the current, we need to differentiate the expression
\beq \label{eq:exres}
  R=-\frac{\rn J'(0)}{\ji}=\frac{\rn \sqrt{2(U(\js)-U(\ji))+\jc^2}}{\ji}
\eeq
with respect to $\ji$. We have:
\barr
  \frac{d R}{d \ji}&=&-\frac{\rn}{\ji}\frac{d J'(0)}{d \ji}+\rn \frac{J'(0)}
  {\ji^2}
\nonumber \\ 
&=&-\frac{\rn}{\ji J'(0)}\frac{d U(\ji)}{d \ji}+\rn \frac{J'(0)}
  {J^2}.
\earr
Substituting $d U(J)/dJ=J''$ and multiplying by a positive quantity
$-\rn J'(0)$, we obtain
\beq
  \frac{d R}{d \ji} \propto \frac{\rn^2 J''(0)}{\ji}-R^2(\ji).
\eeq
The above derivative is definitely positive at $\ji\approx \j0$, since,
as we saw above,  the Cooper pair channel 
gets blocked, and the resistance of the system rises abruptly as $\ji$ 
approaches $\j0$. 
Hence it is only left to determine the dependence on $\ji$ for $\ji<<\j0$. For
this case
we can neglect the quasi-particle contribution and  use the 
expression (\ref{eq:secder}) for $J''(0)$. After a minor manipulation we 
obtain
\beq
  \frac{d R}{d \ji}\propto \frac{\rn \r_0 (1-r^2)}{1-\sqrt{r^2+\x_{in}^2(1-r^2)}}-
  R^2(\ji),
\eeq
where $\x_{in}=J_{in}/J_0$.

\begin{figure}
   \centerline{\psfig{figure=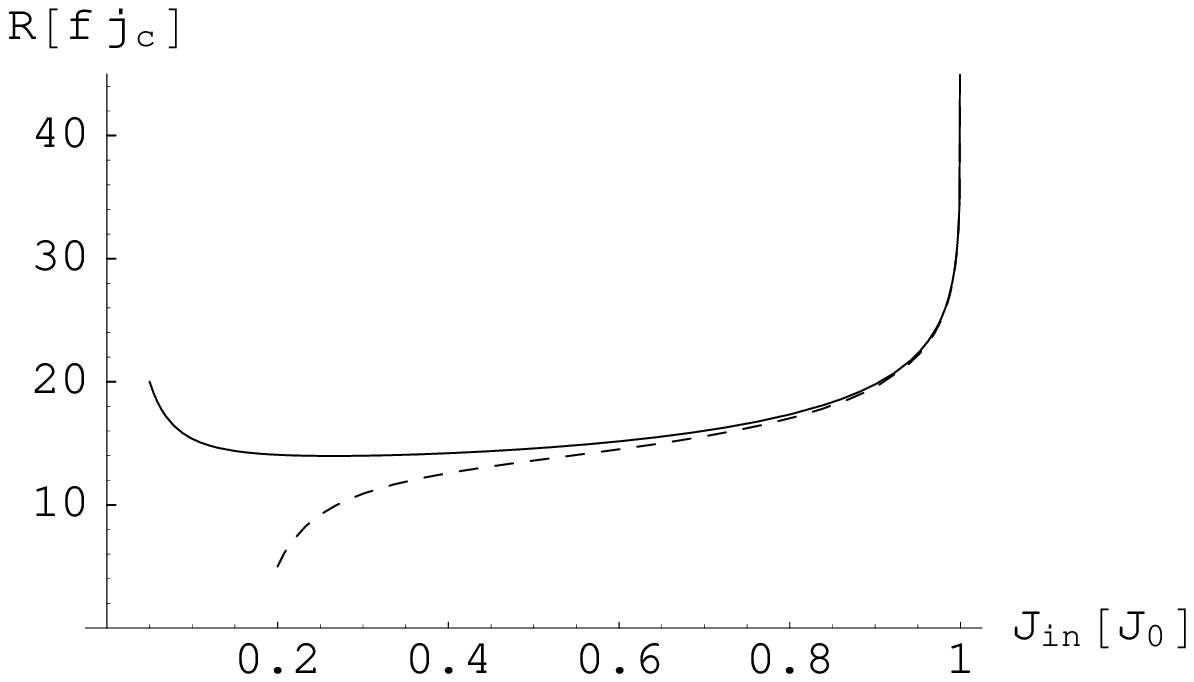,width=8.5cm}}
   \caption{Sample resistance $R(\ji)$ for $\rc<\rco$ (dashed line) and
for $\rc>\rco$ (solid line). }
   \label{rescurr}
\end{figure}

The sign of this expression determines whether the resistance increases or
decreases with the injected current $\ji$. It is easily verified that 
this expression 
is increasing with $\x_{in}$ (i.e. with $\ji$). Hence it is enough to determine
the sign at the smallest current at which the model is applicable, 
$\ji=\js$:  if it is positive,
then the resistance increases monotonically with the current, while if it
is negative, the resistance first decreases and then starts to grow as
the current becomes large enough, c.f. Fig.~\ref{rescurr}. 
Substituting $\x_{in}=
\x_{out}$ and $R(\js)=\rc$, we get:
\end{multicols}
\top{-3cm}
\beq
  \frac{d R}{d \ji} \Big|_{\ji=\js}\propto
 \frac{\rn \r_0(1-r^2)}{1-
  \sqrt{r^2+(\jc f/\rc)^2(1-r^2)}}-\rc^2.
\eeq
The result is a decreasing function of $\rc$. It is positive for small
$\rc$ (which should be larger than $f \jc$ in order to satisfy $\js<\j0$), 
negative for large $\rc$ and vanishes at $\rc=\rco$, given by
\beq
  \rco=
\sqrt{\rn \r_0+\jc^2 f^2/2+\sqrt{(\rn \r_0+\jc^2 f^2/2)^2-\rn^2 
  \r_0^2(1-r^2)}}.
\eeq
\bottom{-2.7cm}
\begin{multicols}{2}
Thus, the dependence of the resistance on the current is controlled by the 
value of $\rc$, as is seen 
in Fig.~\ref{rescurr}. For $f \jc<\rc<\rco$
the resistance increases monotonically with the current. But if
$\rc>\rco$, the resistance decreases for small currents $\js<\ji<J_1$,
where $J_1$ is the solution of an equation
\beq
  \frac{\rn \r_0(1-r^2)}{1-\sqrt{r^2+(J_1/\j0)^2(1-r^2)}}=R^2(J_1).
\eeq

The physical explanation for this behavior is that when the current is 
increased, the 
interface is pushed downwards, increasing the thickness of the upper 
(dissipative)
layer. If the interface is highly conducting (small $\rc$), most of the 
current is shunted through the lower (dissipationless) part of the
system, so the increase in the upper layer thickness increases the 
resistance of the system. However, if the interface is almost insulating (large
$\rc$), most of the current flows through the upper part, and by increasing 
its thickness the
resistance of the system is decreased. Of course, at large enough currents
the rapid increase of $\rz$ due to shear has dominant effect, so the 
resistance increases anyway. As we show below, the relevant case is large
$\rc$, when a re-entrant behavior as a function
of the temperature takes place. Hence, below $\td$ an increase
in the current influences the system in two {\it opposite} ways: it tends to
decrease the resistance by moving the interface downwards;  
while through the effect of shear it tends to increase it. Also, we see 
that a strong increasing dependence of the resistance on the 

\noindent current appears only when $\ji \approx \j0$, this being true 
both above and below $\td$. 

The difference between the results given here and the qualitative arguments of 
Ref.~[\onlinecite{khay}] may be understood in the following qualitative way. Suppose
that a current $\ji$ flows into the system and generates a current profile $J(z)$ with
an interface at $z=d$. When $\ji$ is slightly increased one may expect the current gradient $\pd_z \jx$
to increase, thus increasing $\rz$, increasing anisotropy and pushing the interface upwards. The 
shear-induced increase in $\rz$ and the motion of the interface both tend then to increase the resistance. 
Our model yields a different picture: as $\ji$ is increased, the interface 
  is shifted downward, thus
 reducing the resistance. The motion of the interface and the shear-induced 
increase of $\rz$ operate
then in opposite directions.

\subsection{Dependence on the intralayer resistivity $\rn$}
For the case when there is no interface in the system, increasing $\rn$ makes
the current distribution more homogeneous, so that $\jx(0)$ and $\jx(d)$
differ less. Put in another way, $J_0\equiv \rn/f$ grows. Because of this, the effects of inter-layer vortex shear become weaker, 
and the vertical
resistivity $\rz$ decreases. Hence the total resistance $R$ is influenced 
by two opposite effects: increase of $\rn$ directly increases $R$, this effect
being dominant at small currents. On the other hand,
through the decrease of $\rz$ it tends to decrease $R$, this effect becoming
dominant at strong currents, when effect of shear is important. Hence
the resistance grows with $\rn$ at small $\ji$, while it decreases with
$\rn$ as $\ji$ approaches $\j0$.

\subsection{Dependence on the critical current $\jc$}
Next we discuss the dependence of the sample resistance on the critical 
current.
We disregard a possible dependence of various parameters (like $\rc$, for 
instance) on $\jc$ and consider only a variation of $R$ due to a shift
in the position of the interface and the subsequent current redistribution.
To determine the sign of the derivative $\frac{\pd R}{\pd \jc}$, 
we use the expression
given by Eq. (\ref{eq:exres}). First, it is easy to see that the sign of the 
derivative is
independent of the value of injected current $\ji$. Hence we can find it 
at $\ji=\js$ (i.e. when the interface is right at the top of the sample).
But when this condition is satisfied, the resistance of the system is
constant and equal to $\rc$. Thus,
\beq
  \frac{d R(\ji=\js)}{d \jc}= \frac{\pd R}{\pd \jc}+ 
   \frac{\pd R}{\pd \ji}
   \bigg{|}_{\ji=\js} 
   \frac{d \js}{d \jc} =0.
\eeq
Then, using Eq. (\ref{eq:jout}), we see that
\beq
  \frac{\pd R}{\pd \jc}=-\frac{\rn}{\rc} \frac{\pd R}{\pd \ji} .
\eeq 

\begin{figure}
   \centerline{\psfig{figure=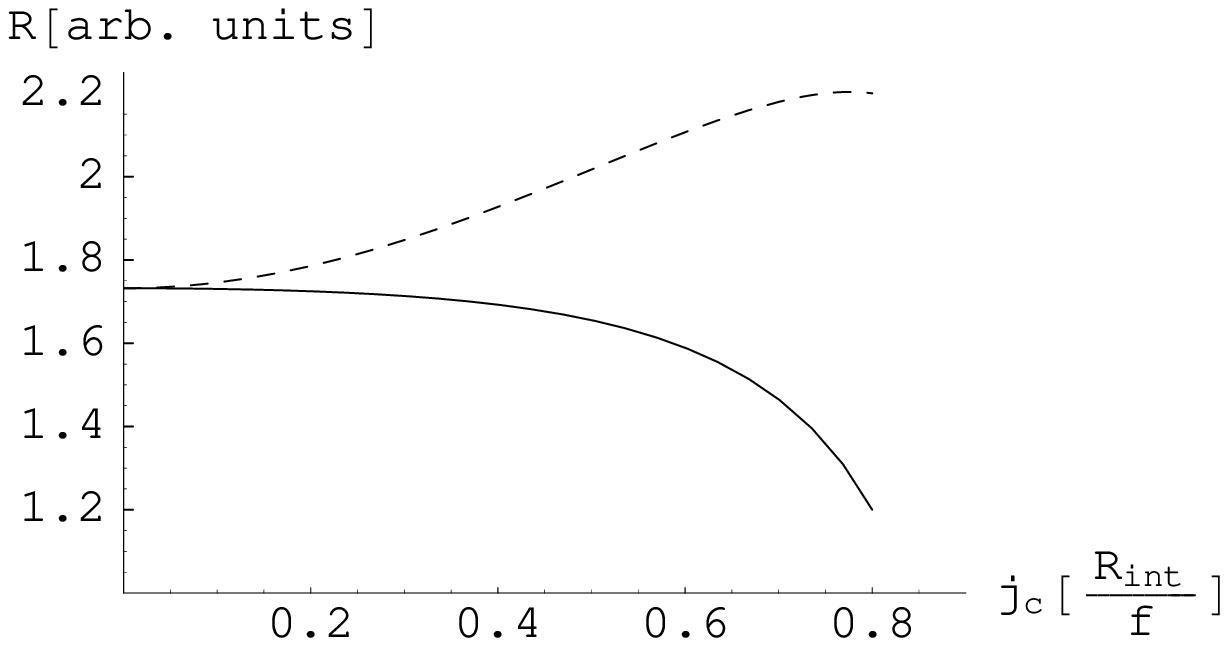,width=8.5cm}}
   \caption{Sample resistance $R(\jc)$ 
at $\ji=0.8\, \j0$  
\newline for
$\rc>\sqrt{\rn(\r_0+\r_1)}$ 
(dashed line) and
for $\rc<\sqrt{\rn(\r_0+\r_1)}$  (solid line). }
   \label{resjc}
\end{figure}

\noindent Consequently, at small $\ji$ the dependence of $R$ on 
$\jc$ is opposite to
its dependence on $\ji$(at larger currents the dependence on $\jc$ remains
of the same type, while the dependence on $\ji$ may change, as was
shown above).  This behavior is natural, as by increasing $\ji$ the interface
is pushed downwards, while increasing $\jc$ it is pushed upwards. Hence
citing the previous results we obtain that $R$ 
increases with $\jc$ for $\rc>\rco$, while it decreases with $\jc$ for 
$\rc<\rco$. Formulated in a different way, this means that $R$ 
increases with $\jc$ when $\jc<j_{c0}$, where $j_{c0}$ is given by
\beq
  j_{c0}=\frac{\rc}{f}\sqrt{\frac{[1-\rn \r_0(1-r^2)/\rc^2]^2-r^2}{1-r^2}}. 
\eeq
In order that $j_{c0}$ be real and positive, $\rc$ has to satisfy $\rc>\sqrt
{\rn(\r_0+\r_1)}$, which is physically plausible, as the interface should
be insulating enough in comparison to the resistive phase in order that 
the rise in its vertical position would increase the sample resistance.
If this condition is not satisfied, or if $\jc>j_{c0}$, $R$ decreases with
$\jc$. Note that when $\rc \rar \infty$ also $j_{c0} \rar \infty$, so that
in this case $R$ increases with $\jc$ for any relevant value of $\jc$. 
The behavior
of the resistance as a function of $\jc$ can be seen in Fig.~\ref{resjc}.
Also, from Eq. (\ref{eq:exres}) we see that 
$\frac{d R}{d \jc}\sim\frac{1}{R}$ (since a square root is differentiated).
Hence at large currents the 
dependence
on $\jc$ becomes weaker. This is in contrast to the dependence on $\ji$,
which becomes very strong as $\ji \rar \j0$. 
All this, of course, is valid when $\jc$ is strong enough that there is an
interface in the system.

\section{Perpendicular resistivity due to parallel current gradient} 
\label{sec:micro}

The interplane transport properties of high-$T_c$ superconductors have been
a subject of intense research over the past decade, both theoretical 
\cite{gray,koshel} and experimental 
\cite{busch,cho,briceno,yurgens,thopart}. This 
transport, being of Josephson nature, is determined by the phase coherence
between the adjacent layers. For a superconductor in a perpendicular
magnetic field, the pancake vortex structure determines the above properties
through the phase distribution. The vortex structure in high-$T_c$ 
superconductors exhibits a rich variety of phenomena, including decoupling,
melting, pinning, Bose glass formation etc., due to thermal fluctuations
\cite{glazman,clem,daemen,blatter}, point defects 
\cite{daemen,blatter,koshel2} or 
columnar defects \cite{blatter,kdv,bmv,morozov,morozov2}. For the 
perpendicular 
resistivity $\r_z$, the microscopic origin of the dissipation is
less obvious than for the in-plane resistivity $\r_x$, where it is understood 
in terms of the Lorentz force, acting on the pancakes. Koshelev \cite{koshel}
proposed a microscopic mechanism for interplane dissipation, in which
the pancake dynamics are shown to influence the interplane conduction, and 
calculated
$\r_z$ for the simplest case of non-interacting pancakes.
Following Ref. \onlinecite{koshel} we analyze a simple microscopic model 
aimed at a derivation of 
a formula for a contribution to the resistivity in 
$\bf{\hat{z}}$
direction (perpendicular to the layers) $\rd_z$
of a superconducting slab due to a gradient in the current in $\bf{\hat{x}}$ 
direction (parallel to the layers). We first derive $\rd_z$ for a 
3-dimensional sample assuming no interactions between the vortex pancakes.
Then we show how the results are modified in presence of interlayer and
intralayer correlations between the pancake positions. Finally, we transform
the 3-D resistivity parameters
into a form 
appropriate for the 1-dimensional  model used in the previous section. That
is, we show how $\rn$, $\r_0$ and $f$ of the macroscopic model are
derived from the resistivities of the 3-dimensional model.

\subsection{Noninteracting pancakes}
We
assume a layered superconductor with noninteracting pancake vortices in it.
The 
vortices
are mobile, and their relative 
diffusive motion provides a mechanism for 
perpendicular resistance. In addition, each layer carries a  different 
current,
causing different drift velocities of vortices in adjacent layers. This
increases the decay of phase correlations in time, thus enhancing the
perpendicular resistance.

We start from the Kubo formula for finite temperatures:
\beq \label{eq:conduc}
  \s = \frac{s j_J^2}{T} \int d \br\,  dt \, \la \sin \df(0,0)\,  \sin 
   \df(\br,t) \ra.
\eeq
Here $s$ is the interplane separation, $j_J$ - the Josephson current and 
$\df$ - the gauge invariant phase difference between neighboring  layers.
We neglect interplane correlations, so that averages like $\la \exp\i\df\ra$
are assumed to be zero and then
\begin{eqnarray}
  &&\la\sin \df(0,0) \sin\df(\br,t)\ra \approx (1/2)Re\la\exp[\i S(\br,t)]\ra,
  \nonumber \\
  &&\mbox{where} \, S(\br,t) \equiv \df(\br,t)-\df(0,0).
\end{eqnarray}
Next we assume Gaussian randomness of $S$, so
that 
\beq 
  \la\exp[\i S(\br,t)]\ra=\exp[-\la S(\br,t)^2\ra/2].
\eeq
Thus we need to calculate the mean square of $S$. We write
\barr
  S(\br,t)= \sum_i&& \pv(\br-\bR_{1,i}(t))-\pv(\br-\bR_{2,i}(t))
\nonumber \\
 &&-\pv(-\bR_{1,i}(0))+\pv(-\bR_{2,i}(0)),
\earr	 
where $\pv(\br)$ is the phase distribution of a single vortex.
Expanding, we write
\barr
  S(\br,t)=\sum_i && [\br-\D\bR_{1,i}(t)]\nf(-\bR_{1,i})
\nonumber \\  && -
[\br-\D \bR_{2,i}(t)]
  \nf(-\bR_{2,i}),
\earr
where $\D\bR(t)\equiv \bR(t)-\bR(0)$ and
\beq \label{eq:gphi}
  \nf(\br)=\frac{{\bf \hat{z}}\times \br}{r^2}.
\eeq
Now we assume that the pancakes in the layers are randomly placed, so that
\begin{eqnarray}
  \la \nf(-\bR_1) \nf(-\bR_2) \ra&=&0 \qquad   \mbox{and} \nonumber \\
  \la \nf(-\bR_{1,i}) \nf(-\bR_{1,j}) \ra&=&\delta_{i,j} 
    \la \nf(-\bR_{1,i})^2 \ra.
\end{eqnarray}
Then the square of a sum breaks into a sum of squares, so that
\barr \label{eq:ssquare}
  \la S(\br,t)^2\ra=\sum_i \,&& \la\left([\br-\D\bR_{1,i}(t)]
   \nf(-\bR_{1,i})\right )^2\ra
\nonumber \\ && +
   \la \left([\br-\D\bR_{2,i}(t)]\nf(-\bR_{2,i})\right)^2\ra.
\earr
Now we can write for each layer
\beq
  \D\bR(t)\approx \bv t+\d\bR(t), 
\eeq
where $\bv$ is the drift velocity of vortices due to the current, and 
$\d\bR(t)$ is the diffusion term. It gives the main contribution at zero 
current gradient, and we will copy it from the Koshelev's article. 
Using the expression (\ref{eq:gphi}) we write
\barr
  \la S(\br,t)^2\ra=&&\sum_i  \,(\br-\bv_1 t)^2 \la\left[ \frac{\bR_{x;1,i}}
  {R^2_{1,i}} \right]^2\ra
\nonumber \\ &&
 +(\br-\bv_2 t)^2 \la\left[ \frac{\bR_{x;2,i}}
  {R^2_{2,i}} \right]^2\ra+ \la S_{diff}(t)^2\ra.
\earr
Now we calculate the averages:
\barr
 \sum_i\, \la \left[\frac{\bR_x}{R^2} \right]^2\ra
 &=&\sum_i\,\frac{1}{2} \la\frac{1}{R^2}\ra=\frac{n}{2}\int \frac{d\bR}{R^2}
\nonumber \\  
 &=&\pi n\ln \frac{R_{max}}{R_{min}},
\earr
where $n$ is the density of the vortices and $R_{min}$ and $R_{max}$ - the
lower and upper cutoff radii.
Substituting this, we obtain 
\end{multicols}
\begin{eqnarray}
 \la S(\br,t)^2\ra&=&[(\br-\bv_1 t)^2+(\br-\bv_2 t)^2]\pi n \ln 
 \frac{R_{max}}{R_{min}}+\la S^2_{diff}(t)\ra \nonumber \\
  &=&[2(\br-\bV t)^2+(\D\bv t)^2/2]\pi n \ln 
 \frac{R_{max}}{R_{min}}+\la S^2_{diff}(t)\ra, 
\end{eqnarray}
where
\beq
  \bV=(\bv_1+\bv_2)/2 \quad \mbox{and} \quad \D\bv=\bv_1-\bv_2.
\eeq

Substituting this result back into Eq. (\ref{eq:conduc}) and using  Koshelev's
result for $S_{diff}$, we obtain:
\begin{eqnarray}
  \s(\D v)&=&\frac{s j^2_J}{2T} \int_{t>0} d\br dt \exp\left(-[(\br-\bV t)^2
   +(\D\bv t)^2/4]
   \pi n \ln \frac{R_{max}}{R_{min}}-2\pi nDt\ln(R_J^2/R_{min}^2)\right) 
   \nonumber \\
   &=&\frac{s j^2_J}{2T}
   \frac{1}{n \ln(R_J/a_0)} \frac{2}{\D v\sqrt{\pi n\ln(R_J/a_0)}} 
   F\left(4D\sqrt{\pi n\ln(R_J/a_0)}/\D v\right), 
\end{eqnarray}
where $D$ is the diffusion constant of pancake motion inside the layers.
We used the Josephson radius $R_J$ for the upper cutoff radius and
the average intervortex spacing $a_0$ - for the lower cutoff.
The function $F(y)$ is defined by
\beq
  F(y)\equiv  \int_0^\infty dx e^{-x^2-2xy}=e^{y^2}\int_y^\infty dx 
  e^{-x^2}
=\frac
  {\sqrt{\pi}}{2}e^{y^2}\left[1-{\rm Erf}(y)\right].
\eeq
This function can be easily approximated for small and large values
of its argument:
\beq
  F(y) \rar \Bigg \lbrace \matrix{ \frac{\sqrt{\pi}}{2}-y & \qquad \mbox{for} 
    \quad y\ll1,
      \cr \frac{1}{2y}-\frac{2}{(2y)^3} & \qquad \mbox{for} \quad y
       \gg1 \cr}.
\eeq
Using this and expressing the vortex velocity difference in terms of parallel
current gradient modulus,
\beq \label{eq:dv}
  \D v=\mu s^2 (\Phi_0/c) \partial_z \jd_x,
\eeq
where the average pancake mobility $\mu$ is connected with the diffusion
constant $D$ by the Einstein relation $D=\mu T$, we obtain for 
the perpendicular resistance 
\beq\label{eq:full}
  \rd_z(\pd_z \jd_x)=\frac{1}{ j^2_J}\frac{2 D s (\Phi_0/c) \pd_z \jd_x  
  [n \ln(R_J/a_0)]^{3/2} \exp(-\pi n \ln(R_J/a_0)[4T/s^2(\Phi_0/c) 
  \pd_z \jd_x]^2) }{1-
  {\rm Erf}[4T\sqrt{\pi n\ln(R_J/a_0)}/s^2 (\Phi_0/c) \pd_z \jd_x]}.
\eeq
Expanding this, we obtain for small current gradients:
\beq
  \rd_z(\pd_z \jd_x)=\frac{T}{s j_J^2}\left(8\pi D[n \ln(R_J/a_0)]^2+
  \frac{1}{4}
  n \ln(R_J/a_0)D[\frac{s^2}{T}\frac{\Phi_0}{c}\pd_z \jd_x]^2\right),
\eeq  
i.e. a parabolic dependence on $\pd_z\jd_x$. On the other hand, for large 
current gradients,
\beq \label{eq:asympt}
  \rd_z(\pd_z \jd_x)=\frac{2D}{s j_J^2}\left([n \ln(R_J/a_0)]^{3/2} s^2
  (\Phi_0/c)
  \pd_z\jd_x+8[n \ln(R_J/a_0)]^2 T\right)
\eeq
i.e., a linear dependence on $\pd_z\jd_x$.
\bottom{-2.7cm}
\bmul

As Eq. (\ref{eq:full}) is not convenient for analytical work, we will use
an approximation of the form $\rd_z(\pd_z \jd_x)=\rd_0+\sqrt{(\rd_1)^2
+(\fd\pd_z \jd_x)^2}$   which gives a correct value at zero current gradient
and the asymptotic behavior at large current gradients. It also approximates
quite well the behavior of $\rd(\pd_z \jd_x)$ in the intermediate range of
current gradients. Comparing the
coefficients, we obtain
\begin{eqnarray} \label{eq:param}
  \rd_0&=&\frac{16DT}{sj_J^2}[n \ln(R_J/a_0)]^2 \nonumber \\
  \rd_1&=&(8\pi-16)\frac{DT}{sj_J^2}[n \ln(R_J/a_0)]^2 \nonumber \\
   \fd&=&\frac{2Ds}{j_J^2}\frac{\Phi_0}{c}[n \ln(R_J/a_0)]^{3/2}.
\end{eqnarray}  
\subsection{Correlations between pancake positions}
Here we demonstrate how the results obtained above are modified  in presence
of inter- and intralayer correlations between pancake positions.

We first consider the effect of interlayer correlations. The presence
of such correlations can be crudely described by regarding pancakes in 
different layers as
tied together into vertical line segments
of length $L_z$, which move as a whole. These segments should
be used instead of independent pancakes of previous subsection. The
phase differences $\df(\br,t)$ and corresponding Josephson currents are
created only at the ends of these segments (more exactly, between layers, 
where one segment ends and another one starts), while the middle
parts of the segments do not contribute to $\df(\br,t)$. This means, that
the effective concentration of vortices is reduced by a factor $L_z/s$.
Next, since each line segment has an increased ``mass'', the mobility 
$\mu$ and the diffusion constant $D$ are now reduced by another factor
$L_z/s$. Finally, the vertical separation between the segments is $L_z$
instead of $s$ for free pancakes. This means that the velocity
difference between the segments due to current gradient is increased
by $L_z/s$. To take into account this and the reduction in the mobility
in Eq.(\ref{eq:dv}), the flux quantum $\Phi_0$ should be multiplied
by $(L_z/s)^2$. This specifies, how the resistivity parameters are
modified in the presence of interlayer correlations.

Next we turn to consider the intralayer correlations. Roughly speaking,
these correlations
cause pancakes in each layer to aggregate in clusters of size $\lxy$,
so that there are $(\lxy/a_0)^2$ pancakes in a cluster. 
Pancakes inside each cluster are ordered, while different clusters move 
independently (actually, there is a hard-core repulsion between them).
Since vortices in the same cluster are not independent, Eq.(\ref
{eq:ssquare}) for the phase correlation square now reads as
\barr
  \la S(\br,t)^2\ra=&&\sum_\a \, \la\left( \sum_{i\in\a} [\br-\D\bR_{1,i}(t)]
   \nf(-\bR_{1,i})\right )^2\ra  \nonumber \\
   && \,+
   \la \left(\sum_{i\in\a} [\br-\D\bR_{2,i}(t)]\nf(-\bR_{2,i})\right)^2\ra,
\earr
where $\a$ is an index of a cluster, while $i$ - of an individual pancake.
For clusters which are far enough away, the differences in the location
of individual pancakes inside the cluster can be neglected. Then each such
cluster gives a contribution to $\la S(\br,t)^2\ra$, which is $(\lxy/a_0)^4$
times larger than a contribution of an individual pancake. On the other hand,
the concentration of the clusters is $n(a_0/\lxy)^2$. To take both effects
into account, we should multiply $n$ by $(\lxy/a_0)^2$ in the final result.
Also, the diffusion constant (and the mobility) of each cluster is
reduced by a factor $(\lxy/a_0)^2$, while the flux quantum $\Phi_0$
should be multiplied by the same factor. Substituting all these prescriptions
into Eq.(\ref{eq:param}), we obtain 
the resistivity parameters in the presence of correlations between pancake 
positions:   
\begin{eqnarray} \label{eq:intparam}
  \rd_0&=&\frac{16DT}{sj_J^2}\frac{(\lxy/a_0)^2}{(L_z/s)^3}
  [n \ln(R_J/a_0)]^2 \nonumber \\
  \rd_1&=&(8\pi-16)\frac{DT}{sj_J^2}\frac{(\lxy/a_0)^2}{(L_z/s)^3}
    [n \ln(R_J/a_0)]^2 \nonumber \\
   \fd&=&\frac{2Ds}{j_J^2}\frac{(\lxy/a_0)^3}{(L_z/s)^{1/2}}
    \frac{\Phi_0}{c}[n \ln(R_J/a_0)]^{3/2}.
\end{eqnarray}  
Here we neglected all changes in the argument of the logarithms.

\subsection{Transformation of the parameters into 1D form}
Now we transform these quantities into a form appropriate for the 
1D macroscopic model. For this, we first define the corresponding fields
and currents from their 3D counterparts (assuming that everything is uniform
in $\uy$ direction):
\barr
  &&\jx(z)=\jd_x(x=L_x/2,z) L_y  \nonumber \\
  &&\jz(z)=\int_0^{L_x/2} dx \,\jd(x,z) L_y \nonumber \\
  &&V(z)=\int_0^{L_x} dx \, \ed_x(x,z)  \nonumber \\
  &&E_z(z)=\ed_z(x=0,z),
\earr
where $L_x$ and $L_y$ are sizes of the sample.
Then, for large current gradients, we use the Ohm's law for the 3D 
sample and average over x:
\barr
  \int_0^{L_x/2} dx\, \ed_x(x,z) &=& \int_0^{L_x/2} dx \, \rd_x \jd_x(x,z) 
  \nonumber \\
  \int_0^{L_x/2} dx \, \ed_z(x,z)&=& \int_0^{L_x/2} dx \, \Big 
  (\rd_0 \jd_z(x,z)
\nonumber \\ &&\, +
  \fd \pd_z\jd_x(x,z) \jd_z(x,z)\Big).
\earr
Defining now the reduced quantities as ratios between the 3D and 1D ones, so 
that
\barr
   \jr_x(x,z)&&\equiv \frac{\jd_x(x,z)}{\jx(z)}  \nonumber \\
   \jr_z(x,z)&&\equiv \frac{\jd_z(x,z)}{\jz(z)}  \nonumber \\
   \er_x(x,z)&&\equiv \frac{\ed_x(x,z)}{V(x,z)}  \nonumber \\
   \er_z(x,z)&&\equiv \frac{\ed_z(x,z)}{E_z(z)},
\earr 
we obtain from the previous equations:
\emul
\barr
   V(z)\int_0^{L_x/2} dx \,\er_x(x,z) &=&\jx(z) \rd_x \int_0^{L_x/2} dx \,
   \jr_x(x,z)  \nonumber \\
   E_z(z) \int_0^{L_x/2} dx \,\er_z(x,z) &=&\jz(z) \rd_0  \int_0^{L_x/2} dx \,
   \jr_z(x,z)  \nonumber \\
   && + \jz(z) \fd \int_0^{L_x/2} dx \, \jr_z(x,z) \pd_z \jx(z) 
   \jr_x(x,z).
\earr 
Then, in order to obtain the equations of the macroscopic model, we make
two assumptions: first, we neglect the derivative $\pd_z \jr_x(x,z)$;
second, we assume that the reduced quantities are not affected by shear
effects, so we calculate them from a linear model with $f=0$.
The parameters of the macroscopic 1D model are then given by
\barr \label{eq:1d}
  \rn &=&\rd_x \int_0^{L_x/2} dx \,\jr_x(x,z) \, {\Bigg /}
  \int_0^{L_x/2} dx \,\er_x(x,z)  \nonumber \\
  \r_0/2 &=&\rd_0 \int_0^{L_x/2} dx \, \jr_z(x,z) \, {\Bigg /}
  \int_0^{L_x/2} dx \,\er_z(x,z)  \nonumber \\
  f/2 &=&\fd \int_0^{L_x/2} dx \,\jr_z(x,z) \jr_x(x,z)
  \, {\Bigg /}
  \int_0^{L_x/2} dx \,\er_z(x,z),
  \earr
\bottom{-2.7cm}
\bmul
\noindent where $\r_0$ and $f$ are divided by $2$, since, as we explained in the 
beginning of the previous section, the perpendicular resistivity of the 
macroscopic model is taken to be $\rz/2$. 

To find the reduced quantities, we need to find the current distribution
in a sample with constant resistivities $\rd_x$ and $\rd_z$. This
amounts to solving the Laplace equation with the boundary conditions
$j_x^0(x=0,z)=j_x^0(x=L_x,z)=j_z^0(x,z=\infty)=0$, and $j_z^0(x,z=0)=0$, 
except two narrow regions near $x=0$ and $x=L_x$, where $j_z^0(x,z=0)$
is, respectively, positive and negative. This describes contacts, attached
to the top of the sample, where the current flows into and out of the system. 
For simplicity we assumed here that the system is infinitely thick in
$\uz$ direction. Choosing an appropriate form for $j_z^0(x,z=0)$, we obtain:
\barr
  j_z^0(x,z)&=&\frac{\sinh{k_0(w+\a z)} \cos{k_0 x}}{\sinh^2{k_0(w+\a z)}
  +\sin^2{k_0 x}}  \nonumber \\
  j_x^0&=&\a \frac{\sin{k_0 x} \cosh{k_0(w+\a z)}}{\sinh^2{k_0(w+\a z)}+
  \sin^2{k_0 x}},
\earr
where $w$ is the width of the contacts ($w\ll L_x$), $\a\equiv
\sqrt{\rd_z/\rd_x}$, and $k_0 \equiv \pi/L_x$. Using this, we calculate
the integrals of the reduced quantities and substitute them into Eq.
(\ref{eq:1d}), thus obtaining

\emul
\top{-3cm}
\barr
  \rn&=&\rd_x \frac{2 L_x}{\pi L_y} \cosh{k_0 (w+\a z)} \log{\coth{k_0 
  (w+\a z)/2}}  \nonumber \\
  \r_0/2&=&\rd_0 \frac{\pi}{L_x L_y} \frac{1}{\sinh{k_0(w+\a z)} 
  \arctan{1/\sinh^2{k_0(w+\a z)}}}  \nonumber \\
  f/2&=&\fd \frac{\pi}{2 L_x L_y^2} \frac{1}{\sinh^2{k_0(w+\a z)} 
  \arctan^2{1/\sinh{k_0(w+\a z)}}}.
\earr
\bottom{-2.5cm}
\bmul
Here a finite $z$ should be taken, so that $k_0 \a z \propto 1$. Then the
hyperbolic functions give factors of order 1, and
the 1D parameters are given by
\barr
  \rn&=&\rd_x \frac{2 L_x}{\pi L_y} \\ \nonumber
  \r_0/2&=&\rd_0 \frac{\pi}{L_x L_y} \\ \nonumber
  f/2&=&\fd \frac{\pi}{2 L_x L_y^2}.
\earr
This establishes a correspondence between 3-dimensional resistivity
parameters and the 1-dimensional ones, which were used in the macroscopic 
model.


\section{Discussion and conclusions}
It is not easy to compare directly predictions of our model with the 
experimental results, since we do not know temperature dependence of various
parameters of the model. Hence we make only qualitative statements based
on robust features of the model.

First, the model predicts that the resistance grows with the current (at least
for not too small currents), and this current non-linearity becomes
very strong as $\ji \rar \j0$. This is consistent with the experimental
result. Using the results
of the microscopic calculation Sec. \ref{sec:micro}, we found that without
correlations between the pancakes $\j0$
is much larger than the relevant $\ji$. However, in presence of correlations 
its value is suppressed by a factor of $(\lxy/a_0)(L_z/s)^{5/2}$, thus making 
its value much closer to $\ji$. If the ratios $\lxy/a_0, L_z/s$ are
assumed to be 10-15, $J_0$ becomes comparable with the experimentally
relevant currents. This
provides an explanation to the experimental fact that the current
non-linearity becomes strong below the depinning transition temperature $T_d$,
where correlations between the pancakes start to build up. 
Next, the model explains the feature of re-entrance, that is, the experimental
observation that below the depinning transition the resistance increases
as the temperature is decreased. According to the model, if the interface
resistance $\rc$ is large enough, the resistance of the system grows with
$\jc$, which naturally starts to grow as the temperature is decreased
below $\td$. Moreover, the model predicts that this rise in the resistance
should be more pronounced for smaller currents, as indeed observed.

Some ingredients are missing from our model.   
First, the model approximates $\rn$ to be 
independent of the intra-layer current. This approximation is presumably good above the 
depinning temperature, but becomes poor below that temperature, where 
intra-layer 
current induces vortex depinning. 
Second, a missing ingredient in our work is a microscopic derivation of the interface 
resistance $\rc$, separating between the resistive and non-resistive parts of 
the sample. The microscopic origin we have in mind is that in the region between
the two phases the pancake mobility is very sensitive to parallel current
variation. Then a small current gradient is enough to create a large
pancake velocity gradient, which would cause a large perpendicular resistance
in that region. Our attempts to provide a microscopic derivation of $\rc$ and 
its temperature 
dependence led us to results that heavily depend on various microscopic 
parameters 
whose values and temperature dependences are not known. We were therefore led 
to leave 
$\rc$ as a phenomenological parameter. 

Altogether, then, our work is able to explain the qualitative features of the 
non-linear
transport observed in  Ref.~\onlinecite{khay} and unravel a unique feature of transport in
superconducting BSCCO samples in perpendicular magnetic field. 

\acknowledgments
We thank E.~Zeldov and V.~Geshkenbein for useful discussions. This work is
supported by the Israel Science Foundation and the Victor Ehrlich chair.

\end{multicols}
\end{document}